\documentclass[twoside,12pt]{article}%
\usepackage{amsmath}
\usepackage{epsfig}
\usepackage{amsfonts}
\usepackage{amssymb}
\usepackage{graphicx}%
\newcommand{\be}{\begin{equation}}
\newcommand{\ee}{\end{equation}}
\newcommand{\bea}{\begin{eqnarray}}
\newcommand{\eea}{\end{eqnarray}}

\begin{document}

\title{Spectral function of a scalar boson coupled to fermions}
\author{Francesco Giacosa$^{(1)}$ and Giuseppe Pagliara$^{(2)}$\\$^{(1)}$\emph{Institut f\"{u}r Theoretische Physik }\\\emph{Johann Wolfgang Goethe - Universit\"{a}t, Max von Laue--Str. 1}\\\emph{D-60438 Frankfurt, Germany}\\$^{(2)}$\emph{Dip.~di Fisica dell'Universit\`a di Ferrara and} \\\emph{INFN Sez.~di Ferrara, Via Saragat 1, I-44100 Ferrara, Italy}}
\maketitle

\begin{abstract}
We present the calculation of the spectral function of an unstable scalar
boson coupled to fermions as resulting from the resummation of the one loop
diagrams in the scalar particle self energy. We work with a large but finite
high-energy cutoff: in this way, the spectral function of the scalar field is
always correctly normalized to unity, independently on the value of the
cutoff. We show that this high energy cutoff affects the Breit-Wigner width of
the unstable particle: the larger the cutoff, the smaller is the width at
fixed coupling. Thus, the existence of a high energy cutoff (alias minimal
length), and for instance the possible opening of new degrees of freedom
beyond that energy scale, could then be in principle proven by measuring, at
lower energy scales, the line shape of the unstable scalar state. Although the
Lagrangian here considered represents only a toy-model, we discuss possible
future extensions of our work which could be relevant for particle physics
phenomenology.

\end{abstract}

\section{Introduction}

The aim of this work is to study the spectral function of a scalar field,
denoted as $H$, coupled via a simple renormalizable Yukawa-type interaction to
a fermion field $\psi$:%
\begin{equation}
\mathcal{L}_{int}=gH\bar{\psi}\psi\text{ .} \label{lint}%
\end{equation}
We assume that the scalar boson $H$ is heavy enough for the decay process
$H\rightarrow\bar{\psi}\psi$ to take place. Thus, $H$ is unstable and has not
a definite mass: a spectral function $d_{H}(x)$ can be obtained as the
imaginary part of the propagator of $H.$ Intuitively, the quantity
$d_{H}(x)dx$ represents the `mass distribution', that is the probability that
the unstable state $H$ has a mass between $x$ and $x+dx$
\cite{salam,Achasov:2004uq,lupo}. While the Breit-Wigner function represents
often a good approximation for $d_{H}(x),$ deviations become evident when a
more advanced treatment of the problem is undertaken. A natural condition
which must be fulfilled is the normalization equation:
\begin{equation}
\int_{0}^{\infty}d_{H}(x)dx=1\text{ ,} \label{norm}%
\end{equation}
which assures the normalization of the probability associated to the mass
distribution, i.e. the normalization of the initial unstable state $H.$ As we
shall see, interesting effects connected to Eq. (\ref{norm}) emerge when
studying this system in detail.

The determination of the propagator of $H$ is a necessary step to obtain its
spectral function, which is proportional to the imaginary part of the
propagator. We consider a fermionic loop which dresses the bare propagator of
$H$ and we perform a resummation of this loop contribution. Simple power
counting shows that the fermionic loop is divergent. Thus, one has to cure the
divergences according to a certain regularization. In this work we shall use
the old-fashioned cutoff regularization\footnote{Of course, more sophisticated
and effective regularization procedures exist (as described later on and in
the Appendix) and are commonly used for calculations. However, the main aim of
this work is conceptual and we thus wish to explicitly keep track of a finite
high-energy scale.}: we thus introduce a finite, albeit large, high energy
scale $\Lambda.$ Namely, we argue that a finite cutoff is better suited to
describe a physical situation, in which high energy contributions are
effectively suppressed when the energy of the particles circulating in the
loop is high enough. Although its precise value, and also the way the high
momenta are suppressed are unknown (a typical choice consists in taking
$\Lambda$ equal to the Planck mass), the finiteness of the cutoff assures that
the condition (\ref{norm}) is always fulfilled. In turn, a logarithmic
dependence on the cutoff cannot be eliminated (by using relations between bare
and dressed parameters): namely, we find that the form of $d_{H}(x)$ is
(weakly) influenced by the precise value of the cutoff. Before discussing this
main property of our results in detail we briefly recall the ideas behind
regularization and renormalization and justify the use of a finite cutoff.

The appearance of divergences in Quantum Field Theory (QFT) plagued its first
stages, up to the development of a successful renormalization program
\cite{peskin,weinberg,itzy,zee}. The first step of the renormalization is the
regularization procedure, in which the divergent integrals appearing at high
orders in perturbation theory are made finite according to a certain
prescription: in the already mentioned cutoff regularization\ the momenta of
the virtual particles are `cut' for high values beyond a certain ultraviolet
(UV) energy scale (the cutoff) $\Lambda$; in the Pauli-Villars approach
particles with a large mass $\Lambda_{PV}$ (which plays the role of the high
energy scale in this scheme) are formally introduced in such a way that the
ultraviolet contributions cancel; in the dimensional regularization the
integrals are evaluated in $4-\epsilon$ dimensions and the divergences appear
as $\epsilon^{-1}$ contributions. The use of a certain regularization scheme
depends on the problem under study. In fact, a regularization can `destroy'
some original symmetries of the Lagrangian, and therefore care is needed. For
instance, a cutoff $\Lambda$ violates gauge invariance (its restoration is
indeed possible, but lengthy \cite{terning,lyubo}), while the Pauli-Villars
and dimensional regularizations preserve it and are therefore usually
preferred in explicit calculations in the framework of gauge theories (though
the Pauli-Villars does not preserve gauge invariance in non-abelian gauge theories).

Once a QFT Lagrangian has been regularized, one can reabsorb the divergences
into the bare parameters of the theory (masses and couplings) plus the
wave-function renormalizations (these steps can be also done by introducing
proper counterterms, which order by order assure that the divergences
disappear). At this point the high energy scale has disappeared from the QFT
and can be formally set to infinity. Each quantity is perfectly finite and
independent on $\Lambda$ (or on $\Lambda_{PV}$ and $\epsilon$). It is well
known that, only for a small subset of QFTs, the renormalizable theories, this
procedure is possible and no divergence (i.e., explicit dependence on the high
energy scale) emerges at higher orders. Indeed, the Lagrangian of the Standard
Model (SM) contains only renormalizable interactions (see e.g. Ref.
\cite{djouadi} and refs. therein).

Non-renormalizable QFTs were regarded in the past as substantially ill-defined
because the high-energy scale does not decouple. Formally, one could introduce
at each order new counterterms, but the price is the need to introduce new
coupling constants at each order. However, it is interesting to stress that
the point of view toward non-renormalizable theories changed in the last
decades. Especially in the framework of Quantum Chromodynamics (QCD), the
development of a variety of QFTs which are not renormalizable was put forward
in order to model nonperturbative QCD phenomena: (i) chiral perturbation
theory is constructed as a theory of the lightest hadronic states (the pions
in its simplest form) \cite{chpt,pich}; the Lagrangian is organized
order-by-order with increasing number of derivatives, which in turn implies an
increasing number of pion momenta in the corresponding Feynman diagrams. The
Lagrangian of chiral perturbation theory is non-renormalizable, but a
successful renormalization program can be carried out order by order. (ii) The
Nambu Jona-Lasinio model is a model of quarks with a quartic, Fermi-like
(non-renormalizable) interaction. A finite QCD-driven cutoff of about $600$
MeV is introduced to correctly describe the vacuum's phenomenology, see for
instance Ref. \cite{klevansky}. (iii) Although the original $\sigma$ model was
renormalizable \cite{oldsm}, modern versions of it are not \cite{dick}.

More in general, nowadays also the SM itself is regarded as an effective model
of a yet-unknown theory which represents its ultraviolet completion. It is
indeed known that at energy larger than the Planck mass gravity effects are
non-negligible. Although a quantum theory of gravity is still not available,
we can conclude that the cutoff of the Standard Model $\Lambda_{SM}$ should be
smaller than the Planck mass, $\Lambda_{SM}\lesssim M_{Planck}$. But this is
an upper limit: $\Lambda_{SM}$ could be much smaller than that, up to the
order of $1$ TeV $=10^{3}$ GeV. It is then plausible to conclude that
$\Lambda_{SM}$ lies in the (quite broad) range $(10^{3},10^{19})$ GeV.

However, as long as the cutoff $\Lambda$ (or, equivalently, $\Lambda_{PV}$) in
a renormalizable theory is finite but much larger than other dimensionful
parameters of the theory, then the results should depend on it at most as
$1/\Lambda$, $1/\Lambda^{2},...$ and are therefore very difficult to be seen
in low-energy processes. (Moreover, such contributions obviously vanish when
the formal limit $\Lambda\rightarrow\infty$ is taken.) Thus, the cutoff is a
physical energy scale, which however does not affect the low-energy behavior
of the theory. This point of view is very well described in\ the QFT book by
Zee \cite{zee}, where it is stressed that the regularization is not only a
mathematical intermediate step but corresponds in some sense to a physical
situation: \textquotedblleft\textit{I emphasize that }$\Lambda$\textit{ should
be thought of as physical, parametrizing our threshold of ignorance, and not
as a mathematical construct. Indeed, physically sensible quantum field
theories should all come with an implicit }$\Lambda$\textit{. If anyone tries
to sell you a field theory claiming that it holds up to arbitrarily high
energies, you should check to see if he sold used cars for a living (pages
146-147 in Ref. \cite{zee}).}\textquotedblright

Having clarified and motivated why we insist on working with a finite cutoff
$\Lambda$, we come back to the purpose of the present work: namely, we aim to
investigate which role plays the cutoff $\Lambda$ on the spectral function
$d_{H}(x)$ of the unstable scalar state $H.$ At a first sight, this seems a
ill-posed question, cause the cutoff $\Lambda$ should not affect, for all the
reasons described above, a physical quantity such as the spectral function.
(In fact, $d_{H}(x)$ can be related -for instance- to fermionic
pair-production process, whose cross section is described in Sec. 2.4.) Quite
surprisingly, we find that this is not the case, and that $d_{H}(x)$ has a
logarithmic dependence on the cutoff: the finite value of $\Lambda$ influences
the width of the peak of the function $d_{H}(x)$. It is then conceivable that
one may pin down the value of the high energy scale $\Lambda$ by studying the
spectral function of the low-energy resonance $H.$ Note, this peculiar
dependence on the cutoff does not take place in superrenormalizable theories,
in which the value of $\Lambda$ does not affect the form of the spectral
function if it is large enough \cite{lupo}. We shall also show that the limit
$x\rightarrow\infty$ and $\Lambda\rightarrow\infty$ do not commute. When
$\Lambda\rightarrow\infty$ is taken first, and consequently the standard
renormalization procedure is applied, no dependence on the cutoff is left, but
a series of inconsistencies emerges: the spectral function is not localized in
the vicinity of the peak, neither for small values of the coupling constant.
This result represents a further hint toward the existence of a finite cutoff.

An immediate application of our formulae can be done in a case which is
reminiscent of the Higgs boson, which has a coupling of the type of Eq.
(\ref{lint}) to fermions. It must be however clearly stressed that with the
simple Lagrangian in Eq. (\ref{lint}) our calculation does not represent a
realistic evaluation of the spectral function of the Higgs boson: namely, no
local gauge invariance is realized in the present simple toy model (and, in
addition, it would also be explicitly broken by the introduction of the
cutoff), only one channel is taken into account and other channels, such as
the four-fermion and the $WW$ ones, are neglected; finally also background
effects are not considered. Thus, the application of our formulae to the case
of the Higgs boson (coupled to only one fermion channel) must be regarded as a
first, simple test to evaluate the possible relevance of the described effect
(the influence of the cutoff). The issue of including finite width effects in
the propagators of the fundamental and unstable particles of the SM is very
complicated and there have been many attempts to solve it (see Ref.
\cite{passarino} and refs. therein). Presently, the complex mass
renormalization scheme \cite{denner,denner2} represents a possible viable
solution, see also Ref. \cite{denner3} for recent developments on the problem
of unitarity in this approach. Again, we do not tackle here the problem of
unstable SM particles, but analyse other (non-perturbative) aspects related to
unstable particles, such as the normalization of their spectral functions,
which cannot be easily investigated within other schemes.

With all these important cautionary comments in mind, it turns anyhow out
that, for the determined Higgs mass of $125$ GeV \cite{lhc}, the Higgs
spectral function is very narrow and thus very well approximated by a simple
Breit-Wigner form. The dependence on the cutoff, although present in
principle, cannot be seen in practice, because its influence on the form of
$d_{H}(x)$ is vanishingly small. On the other hand it is conceivable that other (pseudo)scalar
resonances beyond the minimal SM exist, which are broad and thus could show a
direct dependence of the cutoff in their spectral function.

The paper is organized as follows: in Sec. 2 we present the model and the
calculation of the self-energy and spectral function. In Sec. 3 we show the
numerical results for some interesting cases and finally in Sec. 4 we draw our
conclusions and possible future developments. A rich appendix is also included
in which we discuss different technicalities and subtle points for the
interested reader.

\section{The model and its implications}

\subsection{The Lagrangian}

We study the following renormalizable Lagrangian in which the scalar particle
$H$ (with bare mass $M_{0,H}$) is coupled to the fermion field $\psi$ (with
mass $m_{f}$):%
\begin{equation}
\mathcal{L}=\frac{1}{2}(\partial_{\mu}H)^{2}-\frac{1}{2}M_{0,H}^{2}H^{2}%
+\bar{\psi}(i\gamma^{\mu}\partial_{\mu}-m_{f})\psi+gH\bar{\psi}\psi\text{ ,}
\label{lag}%
\end{equation}
where $g$ is the dimensionless coupling constant. Thus, the Lagrangian
describes a simple Yukawa interaction of a massive fermion with a massive
scalar boson.

\subsection{Decay width}

As a first step we evaluate the tree-level decay width for the process
$H\rightarrow\bar{\psi}\psi.$ For future purposes we evaluate it for the
arbitrary mass $x$ of the particle $H$:
\begin{equation}
\Gamma_{H\rightarrow\bar{\psi}\psi}^{\text{t-l}}(x)=\frac{\sqrt{\frac{x^{2}%
}{4}-m_{f}^{2}}}{8\pi x^{2}}(4m_{f}^{2})\sum_{\alpha,\beta}\left\vert
\mathcal{M}^{\alpha\beta}\right\vert ^{2}\theta(x-2m_{f}) \label{gtl}%
\end{equation}
where the amplitude reads:
\begin{equation}
-i\mathcal{M}^{\alpha\beta}=-ig\bar{u}^{(\alpha)}(\vec{k}_{1})v^{(\beta)}%
(\vec{k}_{2})
\end{equation}
Following the usual steps (details in Appendix A.1) the tree-level decay width
$\Gamma_{H\rightarrow\bar{\psi}\psi}(x)$ as function of the (running) mass $x$
reads:%
\begin{equation}
\Gamma_{H\rightarrow\bar{\psi}\psi}^{\text{t-l}}(x)=\frac{\left(  \frac{x^{2}%
}{4}-m_{f}^{2}\right)  ^{3/2}}{\pi x^{2}}g^{2}\theta(x-2m_{f})\text{ .}%
\end{equation}
Naively, the on-shell tree-level decay width is evaluated by setting
$x=M_{0,H}.$ However, care is needed because it is a well known fact that the
mass of the $H$ field is modified by loop corrections. In particular, we will
see in the subsections 2.3 and 2.4 that:%
\begin{equation}
M_{0,H}\overset{\text{loops}}{\rightarrow}M_{H}<M_{0,H}\text{ ,}%
\end{equation}
i.e. the loops reduce the mass. The numerical value of the tree-level decay
width is obtained by evaluation the tree-level decay function at the dressed
mass $M_{H}$ (and not at the bare mass $M_{0,H}$): $\Gamma_{H\rightarrow
\bar{\psi}\psi}^{\text{t-l}}(x=M_{H}).$ This procedure is a consequence of 
renormalization: the mass counterterm added to the 
Lagrangian automatically leads to a tree level decay width computed
at the dressed mass which is the physical and thus measurable mass, see Appendix
A.2.1 for details.

The spectral function, to be studied in details later, can be approximated by
the following schematic behavior:
\begin{equation}
d_{H}^{\text{appr}}(x)\simeq\frac{2x}{\pi}\frac{x\Gamma_{H\rightarrow\bar
{\psi}\psi}^{\text{t-l}}(x)}{(x^{2}-M_{H}^{2})^{2}+x^{2}\left(  \Gamma
_{H\rightarrow\bar{\psi}\psi}^{\text{t-l}}(x)\right)  ^{2}}\text{ ,}
\label{dhappr}%
\end{equation}
where the real part and cutoff-effects in the imaginary part of the loop have
been neglected. For large $x$, the approximate asymptotic behavior
$d_{H}^{\text{appr}}(x)\sim1/x$ holds because the decay function
$\Gamma_{H\rightarrow\bar{\psi}\psi}^{\text{t-l}}(x)$ scales as $\Gamma
_{H\rightarrow\bar{\psi}\psi}^{\text{t-l}}(x)\sim x$. Such a spectral function
is clearly non normalized. We shall elaborate on this issue more in detail in
the following, where we will show that the presence of a cutoff (no matter how
large) assures that the correct normalization $\int_{0}^{\infty}d_{H}(x)dx=1$
is obtained.

\subsection{The fermionic loop}

The scalar state $H$ is dressed by fermion loops. The contribution of one
fermion loop is easily evaluated by using the Feynman rules:%
\begin{equation}
\Sigma(p)=i\int\frac{d^{4}q}{(2\pi)^{4}}Tr\left[  \Delta_{f}(q+p/2)\Delta
_{f}(q-p/2)\right]  \text{ ,} \label{sigma}%
\end{equation}
where the fermion propagator reads
\begin{equation}
\Delta_{f}(q)=\frac{1}{\gamma^{\mu}q_{\mu}-m_{f}+i\varepsilon}\text{ .}%
\end{equation}

The integral in Eq. (\ref{sigma}) is quadratically divergent. It must be
therefore regularized; for the reasons described in the Introduction we use
here a regularization function $\phi_{\Lambda}(p,q),$ which depends on the
cutoff $\Lambda$:%
\begin{equation}
\Sigma(p)=i\int\frac{d^{4}q}{(2\pi)^{4}}Tr\left[  \Delta_{f}(q+p/2)\Delta
_{f}(q-p/2)\right]  \phi_{\Lambda}^{2}(p,q)\text{ .}%
\end{equation}

Upon one-loop resummation the propagator of $H$ takes the form%
\begin{equation}
\Delta_{H}(p)=\frac{1}{p^{2}-M_{0,H}^{2}+g^{2}\Sigma(p)+i\varepsilon}\text{ ,}%
\end{equation}
where the loop $\Sigma(p)$ can be rewritten in the following way:%
\begin{equation}
\Sigma(p)=i\int\frac{d^{4}q}{(2\pi)^{4}}\frac{\mathrm{Tr}\left[  \left[
\gamma^{\mu}(q_{\mu}+\frac{1}{2}p_{\mu})+m_{f}\right]  \left[  \gamma^{\nu
}(q_{\nu}-\frac{1}{2}p_{\nu})+m_{f}\right]  \right]  }{\left[  (q+p/2)^{2}%
-m_{f}^{2}+i\varepsilon\right]  \left[  (q-p/2)^{2}-m_{f}^{2}+i\varepsilon
\right]  }\phi_{\Lambda}^{2}(p,q)\text{ .} \label{integral}%
\end{equation}
For what concerns $\phi_{\Lambda}(p,q)$ we make here the following assumption:%
\begin{equation}
\phi_{\Lambda}(p,q)=f_{\Lambda}\left(  \frac{q^{2}p^{2}-(q\cdot p)^{2}}{p^{2}%
}\right)  \text{ .}%
\end{equation}
Notice that the function is expressed in terms of scalar products of
four-vectors and it is thus manifestly covariant. On a practical level we use
the following form for $f_{\Lambda}(\eta)$:%
\begin{equation}
f_{\Lambda}(\eta)=\theta(\eta+\Lambda^{2})\text{ ,} \label{cutofffunct}%
\end{equation}
where $\Lambda$ is a cutoff, see Appendix A.2.1 for more technical details.
The choice in Eq. (\ref{cutofffunct}) is simple and allows for an analytic
presentation of many formulae. However, one could have used smooth and more
complicated cutoff functions, see for instance Refs. \cite{lupo} and refs.
therein. Only small numerical changes would be found but no conceptual changes
would follow.

The trace in the integral (\ref{integral}) reads:
\begin{align}
&  \mathrm{Tr}\left[  \left[  \gamma^{\mu}\left(  q_{\mu}+\frac{1}{2}p_{\mu
}\right)  +m_{f}\right]  \left[  \gamma^{\nu}\left(  q_{\nu}-\frac{1}{2}%
p_{\nu}\right)  +m_{f}\right]  \right] \nonumber\\
&  =4\left(  q+\frac{p}{2}\right)  \cdot\left(  q-\frac{p}{2}\right)
+4m_{f}^{2}=4\left(  q^{2}-\frac{p^{2}}{4}+m_{f}^{2}\right)  \text{ .}%
\end{align}
Then, working in the reference frame of the particle $H$, for which
$p=(x,0)\rightarrow p^{2}=x^{2}$, and performing the integral over $q^{0}$ by
utilizing the residues calculus, one finds:%
\begin{equation}
\Sigma(x)=\int\frac{d^{3}q}{(2\pi)^{3}}\frac{4f_{\Lambda}^{2}(-\mathbf{q}%
^{2})}{2x\sqrt{\mathbf{q}^{2}+m_{f}^{2}}}\frac{4x\mathbf{q}^{2}}{\left[
4(\mathbf{q}^{2}+m_{f}^{2})-x^{2}+i\varepsilon\right]  }\text{ ,}%
\end{equation}
where we have taken into account that, in the rest frame of $H$, one has
$\phi_{\Lambda}(p,q)=f_{\Lambda}(-\mathbf{q}^{2})$ (i.e., no explicit
dependence on $q^{0}$ is present). Introducing the variable $w$ defined as
$w^{2}=\mathbf{q}^{2}$ we rewrite the loop as:%
\begin{equation}
\Sigma(x)=\frac{1}{2\pi^{2}}\int_{0}^{\infty}dw\frac{8f_{\Lambda}^{2}(-w^{2}%
)}{\sqrt{w^{2}+m_{f}^{2}}}\frac{w^{4}}{\left(  4(w^{2}+m_{f}^{2}%
)-x^{2}+i\varepsilon\right)  }\text{ .} \label{lupo}%
\end{equation}
The quadratic divergence of the loop is again clear. The validity of the
optical theorem%
\begin{equation}
x\Gamma_{H\rightarrow\bar{\psi}\psi}^{\text{t-l}}(x)f_{\Lambda}\left(
-\sqrt{\frac{x^{2}}{4}-m_{f}^{2}}\right)  =g^{2}\operatorname{Im}%
\Sigma(x)\text{ }%
\end{equation}
can be easily verified from Eq. (\ref{lupo}) by an explicit calculation of the
imaginary part.

Strictly speaking, the `correct' tree-level decay width, including the effect
of the cutoff function, is given by%
\begin{equation}
\Gamma_{H\rightarrow\bar{\psi}\psi}^{\text{t-l,correct}}(x)=\Gamma
_{H\rightarrow\bar{\psi}\psi}^{\text{t-l}}(x)f_{\Lambda}\left(  -\sqrt
{\frac{x^{2}}{4}-m_{f}^{2}}\right)
\end{equation}
where the vertex-function directly enters into the expression. This result can
be achieved by using a nonlocal Lagrangian and its corresponding Feynman
rules, see Appendix A.2.1 and Refs. \cite{lupo,lyubo,nonlocal}.

In this work we make the choice in Eq. (\ref{cutofffunct}), for which an
analytic form of the loop is obtained as:
\begin{align}
\Sigma(x)  &  =\frac{-1}{4\pi^{2}x}\left\{  \left(  x^{2}-4m_{f}^{2}\right)
^{3/2}\text{arctanh}\left[  \frac{\Lambda x}{\sqrt{\left(  \Lambda^{2}%
+m_{f}^{2}\right)  \left(  x^{2}-4m_{f}^{2}\right)  }}\right]  \right.
\nonumber\\
&  \left.  +x\left[  -2\Lambda\sqrt{\Lambda^{2}+m_{f}^{2}}+(6m_{f}^{2}%
-x^{2})\ln\left(  \frac{\Lambda+\sqrt{\Lambda^{2}+m_{f}^{2}}}{m_{f}}\right)
\right]  \right\}  \text{ .} \label{lupoexpl}%
\end{align}
We shall use the previous form for analytic and numerical calculations. Note
that, as long as $x$ fulfills the inequality%
\begin{equation}
-\sqrt{\frac{x^{2}}{4}-m_{f}^{2}}+\Lambda>0\rightarrow x<2\sqrt{\Lambda
^{2}+m_{f}^{2}}\lesssim2\Lambda\label{constr}%
\end{equation}
one has for the adopted choice of the vertex function that $\Gamma
_{H\rightarrow\bar{\psi}\psi}^{\text{t-l,correct}}(x)=\Gamma_{H\rightarrow
\bar{\psi}\psi}^{\text{t-l}}(x)$. However, for $x>2\sqrt{\Lambda^{2}+m_{f}%
^{2}}$ the correct tree-level decay function vanishes. Thus, for large values
of the cutoff the equality $\Gamma_{H\rightarrow\bar{\psi}\psi}%
^{\text{t-l,correct}}(x)=\Gamma_{H\rightarrow\bar{\psi}\psi}^{\text{t-l}}(x)$
holds for a very wide energy range. (Note, using a smooth cutoff function the
strict equality $\Gamma_{H\rightarrow\bar{\psi}\psi}^{\text{t-l,correct}%
}(x)=\Gamma_{H\rightarrow\bar{\psi}\psi}^{\text{t-l}}(x)$ would hold only
approximately in a wide energy region.)

\begin{figure}[ptb]
\begin{centering}
\epsfig{file=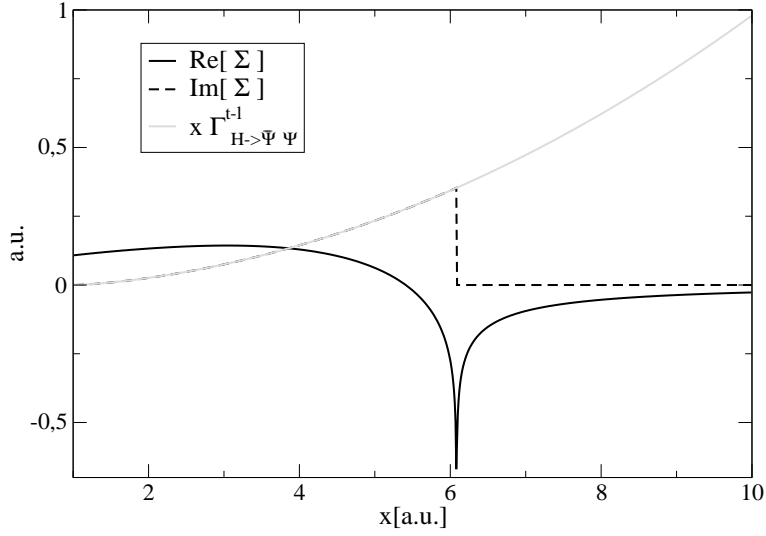,height=10cm,width=7cm,angle=-90}
\caption{Real and imaginary part of the self energy. The quantity $x \Gamma^{t-l}_{H\rightarrow \bar\Psi \Psi}(x)$ is also shown
to confirm the validity of the optical theorem. Because of the cutoff, introduced for the sake of having a correct normalization of the
spectral function, the imaginary part of the self energy vanishes at high energies. Here
$M_H=1.5$ and $\Lambda=3$, where the unit $2m=1$ has been used.}
\end{centering}
\end{figure}

In Fig. 1 we show the results for the real and imaginary part of the self
energy for the following choice of the free parameters: $2m_{f}\equiv1$ sets
an arbitrary energy unit, $M_{H}=1.5$ and $\Lambda=3$. Notice that the
imaginary part vanishes for values of the energy larger than $2\sqrt
{\Lambda^{2}+m_{f}^{2}}$ as explained before. The gray line represents the
quantity $x\Gamma_{H\rightarrow\bar{\psi}\psi}^{\text{t-l}}(x)$ which, due to
the optical theorem, is equal to the imaginary part of the self energy up to
$2\sqrt{\Lambda^{2}+m_{f}^{2}}$. The real part becomes also very small for
$x\gtrsim2\Lambda$: this is a crucial property to show the correct
normalization, as it is presented in the next subsection and in the Appendix A.3.

A closer inspection of the loop expression (\ref{lupoexpl}) shows that
constant terms can still be reabsorbed in the bare mass $M_{0,H}$ and have
therefore no physical consequences; however, this does not hold for the the
term proportional to $x^{2}\ln\Lambda,$ which is responsible for a mixing of
two, in principle well separated energy scales, i.e. $x$ (which is the
invariant mass in a scattering experiment, see the following discussion) and
the cutoff $\Lambda$. Thus, in our scheme there is a logarithmic dependence of
the cutoff in the loop formula and, consequently, on the form of the spectral
function $d_{H}(x)$. This is indeed crucial for our results because this
dependence on the high energy scale $\Lambda$ does not decouple. Indeed, in
Ref. \cite{zee} it was stated that the physical results (such as scattering
lengths) do depend on the cutoff in a power-suppressed form $1/\Lambda,$
$1/\Lambda^{2},...$. The new point here is that we find a dependence of the
cutoff which is logarithmic and not power-like suppressed. Of course, a
logarithmic dependence is weak, but can lead to interesting phenomena, as we
shall see in Sec. 3. \begin{figure}[ptb]
\begin{centering}
\epsfig{file=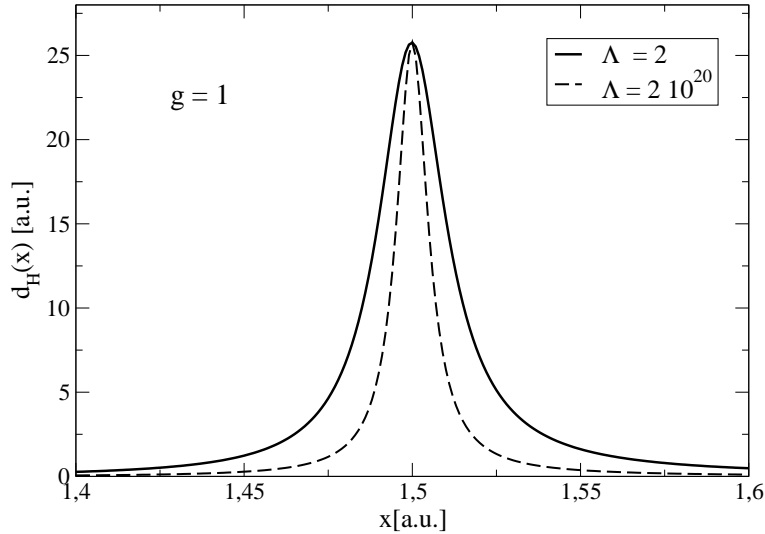,height=10cm,width=7cm,angle=-90}
\caption{Spectral functions at fixed coupling $g=1$ and for two very different
choices of the cutoff value. Notice the narrowing of the spectral function as the cutoff
is increased. $M_H$ as in Fig. 1.}
\end{centering}
\end{figure}

\subsection{The spectral function and its normalization}

The spectral function (or mass distribution) of the scalar field $H$, denoted
as $d_{H}(x)$, is defined as:
\begin{equation}
d_{H}(x)=\lim_{\varepsilon\rightarrow0^{+}}\frac{2x}{\pi}\left\vert
\operatorname{Im}\Delta_{H}(x)\right\vert \text{ .}%
\end{equation}
Explicitly:%

\begin{equation}
d_{H}(x)=\lim_{\varepsilon\rightarrow0^{+}}\frac{2x}{\pi}\frac{g^{2}\left\vert
\operatorname{Im}\Sigma(x)\right\vert }{\left[  x^{2}-M_{0,H}^{2}%
+g^{2}\operatorname{Re}\Sigma(x)\right]  ^{2}+\left[  g^{2}\operatorname{Im}%
\Sigma(x)\right]  ^{2}+i\varepsilon}\text{ .}%
\end{equation}
We define the nominal mass $M_{H}$ of the resonance $H$ as the zero of the
real part of the propagator's denominator\footnote{See Ref. \cite{e38} for a
related study in which, instead of the nominal mass, the pole on the
II-Riemann sheet is investigated.}:%
\begin{equation}
x^{2}-M_{0,H}^{2}+g^{2}\operatorname{Re}\Sigma(x)=0\rightarrow x=M_{H}.
\label{nommass}%
\end{equation}
In general, the quantum loop generates a negative contribution, therefore
$M_{H}<M_{0,H}$.

We assume here that $\operatorname{Im}\Sigma(x=M_{H})\neq0$ (we are thus above
threshold: $M_{H}>2m$ and the decay channel $H\rightarrow\bar{\psi}\psi$ is
open). Then, the limit can be easily performed:%
\begin{equation}
d_{H}(x)=\frac{2x}{\pi}\frac{g^{2}\left\vert \operatorname{Im}\Sigma
(x)\right\vert }{\left[  x^{2}-M_{0,H}^{2}+g^{2}\operatorname{Re}%
\Sigma(x)\right]  ^{2}+\left[  g^{2}\operatorname{Im}\Sigma(x)\right]  ^{2}%
}\text{ .} \label{dh}%
\end{equation}
In the limit $g\rightarrow0$ the expected result $d_{H}(x)=\delta(x-M_{0,H}$)
is obtained. One can compare Eq. (\ref{dh}) with the previously approximate
version in Eq. (\ref{dhappr}): besides the real part, which is neglected in
the approximate form, the two expressions coincide in virtue of the optical
theorem for $x<2\sqrt{\Lambda^{2}+m_{f}^{2}}$.

We now turn to the normalization of $d_{H}(x),$ i.e. to the validity of the
equation:
\begin{equation}
\int_{0}^{\infty}\mathrm{dx}d_{H}(x)=1\text{ .}%
\end{equation}
To this end we recall that the propagator $\Delta_{H}(x)$ can be expressed via
the so-called K\"{a}llen-Lehman representation \cite{itzy}:%
\begin{equation}
\Delta_{H}(x)=\int_{0}^{\infty}\mathrm{dy}\frac{d_{H}(y)}{x^{2}-y^{2}%
+i\varepsilon}\text{ ,} \label{kl}%
\end{equation}
which intuitively corresponds to expressing the full propagator as the `sum'
of free propagators of the form $\left[  x^{2}-y^{2}+i\varepsilon\right]
^{-1}$, each of them weighted with the mass distribution $d_{H}(y).$ The
physical interpretation of $d_{H}(y)\mathrm{dy}$ as the probability that the
unstable state $H$ has a mass between $y$ and $y+\mathrm{dy}$ is evident. When
considering the limit $x\rightarrow\infty$ the propagator can be approximated
as
\begin{equation}
\Delta_{H}(x)\overset{x^{2}\rightarrow\infty}{\simeq}\frac{1}{x^{2}}
\label{proplargep}%
\end{equation}
\emph{provided that} a finite (no matter how large) cutoff $\Lambda$ is
employed. In fact, in the case of a finite cutoff one has that $g^{2}%
\operatorname{Im}\Sigma(x^{2})=x\Gamma_{H\rightarrow\bar{\psi}\psi
}^{\text{t-l,correct}}=0$ for $x>2\sqrt{\Lambda^{2}+m_{f}^{2}},$ see the
previous Section, and also the real part of $\Sigma(x)$ goes rapidly to zero
for $x\rightarrow\infty$. When Eq. (\ref{proplargep}) holds (i.e. the cutoff
is finite), Eq. (\ref{kl}) reduces to%
\begin{equation}
\Delta_{H}(x\rightarrow\infty)=\frac{1}{x^{2}}=\int_{0}^{\infty}%
\mathrm{dy}\frac{d_{H}(y)}{x^{2}}\rightarrow\int_{0}^{\infty}\mathrm{dy}%
d_{H}(y)=1\text{ .}%
\end{equation}
(See also the Appendix A.3 for a rigorous proof and for its extension to the
case of a generic cutoff function, as long it vanishes sufficiently fast for
large values of $x\gtrsim\Lambda$.)

The case of a large but finite cutoff corresponds to realistic cases. In fact,
the cutoff within a `fundamental renormalizable theory' signalizes the own
breaking of the theory and its numerical value is typically much larger than
the other energy scales of the theory (such as the Planck mass). Moreover, we
have shown that the finite cutoff (independently on its value) assures that
the mass distribution is already correctly normalized to unity. There is no
need of a field strength renormalization in this framework; see next Section
for numerical examples.

Some important points need to be discussed.

\bigskip

\emph{Standard renormalization treatment:} It is possible, using the standard
procedure, to remove each dependence on the cutoff. However, we shall show
here and in the Appendix A.4 that inconsistencies arise. The first step
consists in choosing a very large cutoff, $\Lambda\gg M_{H}$, which allows to
simplify the formula (\ref{lupoexpl}) for $x\ll\Lambda$ as follows:
\begin{equation}
\tilde{\Sigma}(x)=-\frac{\left(  x^{2}-4m_{f}^{2}\right)  ^{3/2}}{4\pi^{2}%
x}\text{arctanh}\left(  \frac{x}{\sqrt{x^{2}-4m_{f}^{2}}}\right)
+\frac{\Lambda^{2}}{2\pi^{2}}+\frac{(x^{2}-6m_{f}^{2})}{4\pi^{2}}\ln\left(
\frac{2\Lambda}{m_{f}}\right)  \text{ } \label{lupoappr}%
\end{equation}
Taking the limit $\Lambda\rightarrow\infty$, the propagator corrections
require a quadratically divergent mass renormalization to reabsorb the term
$\sim\Lambda^{2}$ and a field strength renormalization to reabsorb the term
$\sim x^{2}\ln\Lambda$ \cite{peskin}. This way can be easily followed in the
case of a stable scalar particle, $M_{H}<2m_{f}$: these two operations
correspond to the conditions that the pole of the propagator occurs at $M_{H}$
and that the residue at the pole is $1$. When $M_{H}>2m_{f}$ (i.e., $H$ is
unstable) this approach can be formally generalized, although it is not
evident which constraint should be imposed to fix the renormalization
constants. We shall discuss the possibilities in Sec. A.4 where we describe in
detail the relevant procedure and formulae.

After the renormalization procedure, no dependence on the cutoff is present
and the limit for large $x$ of the loop function reads $\Sigma(x)\sim x^{2}\ln
x$ (and does not vanish as Eq. (\ref{lupoexpl}) does.) As a consequence, the
propagator in this case has a different scaling than the one in Eq.
(\ref{proplargep}): $\Delta_{H}(x)\overset{x^{2}\rightarrow\infty}{\simeq
}1/(x^{2}\ln x)$.\ Thus, the limits $x\rightarrow\infty$ and $\Lambda
\rightarrow\infty$ do \emph{not} commute: this is indeed the main origin of
the (very) different results obtained in our framework and the ones of the
standard renormalization. Moreover, due to the different scaling law, the
correct normalization of the spectral function is not anymore guaranteed.
Although the integral of the spectral function is still (slowly) convergent,
one finds rather unphysical results: only a minimal part of the normalization
of the spectral function is located in the vicinity of the peak, implying that
the probability to excite the resonance at energies close to its nominal mass
(corresponding to the position of the peak) is very small. Moreover, also the
dependence of the normalization on the coupling constant $g$ turns out to be
unexpected: the smaller $g$, the smaller is the probability that the unstable
state $H$ has a mass close to the peak. We regard these properties as unphysical.

\bigskip

\emph{Other regularization procedures: } For completeness we have performed in
the Appendix, Secs. A.2.2 and A.2.3, the calculation of the loop in the
Pauli-Villars and the dimensional regularization schemes. In both cases, as
expected, the imaginary part coincides with that obtained in the cutoff scheme
if it is sent to infinity, in agreement with the optical theorem. In the
Pauli-Villars a very similar expression for $d_{H}(x)$ with a finite cutoff
$\Lambda$ is obtained in the vicinity of the peak (including a logarithmic
term of the type $x^{2}\ln\Lambda_{PV}$, as long as the corresponding cutoff
$\Lambda_{PV}$ is finite). However, the Pauli-Villars approach breaks
unitarity already at the level of the Lagrangian and for $\Lambda
_{PV}\rightarrow\infty$ the very same problems described above and in Appendix
A.4 arise. In the dimensional regularization, instead of the term proportional
to $x^{2}\ln\Lambda$, a similar term proportional to $x^{2}/\epsilon$ is
present: removing the latter is also completely equivalent to the case
described above and in Appendix A.4.

\bigskip

\emph{ Gedanken experiment:} One may ask to which extent the spectral function
$d_{H}(x)$ is a physical quantity. To show that $d_{H}(x)$ can be considered
such, we present a `Gedanken experiment', in which the introduced mass
distribution $d_{H}(x)$ directly enters into the form of the total cross
section. To this end, let us consider a (for simplicity massless) scalar field
$\varphi$ which is coupled to the scalar field $H$ via the following
interaction term:%
\begin{equation}
\mathcal{L}_{int,\varphi}=\lambda H\varphi^{2}\text{ .}%
\end{equation}
Writing $x=\sqrt{s}$, the cross section for the fermionic pair
production process $\varphi\varphi\longrightarrow\bar{\psi}\psi$ takes the form%

\begin{equation}
\sigma(x)=\frac{\pi}{2}\frac{\lambda^{2}}{x^{3}}d_{H}(x) \label{cs}%
\end{equation}
which shows that the mass distribution $d_{H}(x)$ directly enters into a
`measurable' quantity. In Ref. \cite{lupo,nonexp,duecan} a somewhat related
`Gedanken experiment' with a decaying particle was described, in which
$d_{H}(x)$ also emerged as a measurable quantity. Indeed, for a nice example
from hadron physics we refer to the radiative decay of the $\phi$ meson
theoretically described with the help of spectral functions in Ref.
\cite{lupoder} and experimentally measured in Ref. \cite{kloe}.

Note that, a new kind of particle is introduced in our
Gedankenexperiment because the virtual state $H$ appears only in the $s$
channel, thus making Eq. (33) valid and simplifying the discussion. In fact,
the $t$ and the $u$ channels do not enter in such a production process. 

In principle there is no restriction on the value of the dimensionful coupling
constant $\lambda$ of Eq. (32). One should include the loops of the bosonic
$\varphi$-field into the evaluation of the propagator and the spectral
function of the field $H.$ Namely, being the interaction in Eq. (32)
superrenormalizable, it does not affect the described influence of the
high-energy cutoff on the spectral function $d_{H}(x)$. However, since here
the bosonic $\varphi$-field is only a mathematical tool of our Gedankenexperiment
for the generation of the virtual state $H$ and its spectral function, 
we assume for simplicity that $\lambda$ is small enough such that the
propagator of the state $H$ is to a very good accuracy determined by the loops
of the fermionic field $\psi$ only. This means that the decay $\Gamma
_{H\rightarrow\varphi\varphi}=\lambda^{2}/(8\pi M_{H}^{{}})$ is assumed to be
much smaller than $\Gamma_{H\rightarrow\bar{\psi}\psi},$ a condition which is
satisfied for $\lambda\ll gM_{H}.$ It should be anyhow stressed that Eq. (33)
will not be used further, it represents just a simple example on how
$d_{H}(x)$ can enter into the expression of a measurable quantity such as a
cross-section.

\section{Narrowing of the width for increasing cutoff: numerical results}

We present now the numerical results for the spectral function of the boson
$H$ coupled to fermions via the Lagrangian (\ref{lag}) using the expression
(\ref{lupoexpl}) with a finite value of the cutoff. For the conceptual purpose
followed here, we do not refer to a particular physical system but present the
results in terms of an energy unit equal to $2m_{f}\equiv1$.

The are five parameters entering in the model: besides the bare parameters
$m_{f},$ $g,$ $M_{0,H}$, there are also the two wave-function renormalization
$Z_{\psi}$ and $Z_{H}$.

\begin{figure}[ptb]
\begin{centering}
\vskip 0.1cm
\epsfig{file=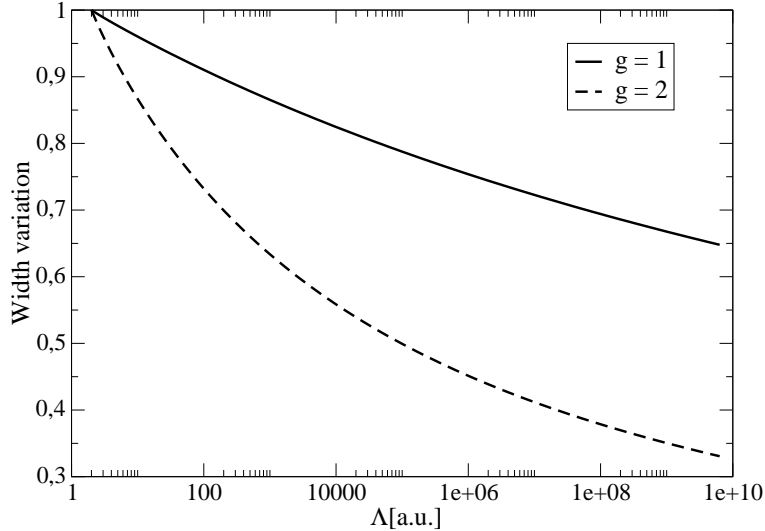,height=10cm,width=7cm,angle=-90}
\caption{Variation of the Breit-Wigner width as a function of the cutoff and for two values
of the coupling. The larger the coupling the faster the width drops for increasing values of the cutoff. }
\end{centering}\end{figure}

In the one-loop study presented here, however, no loop corrections to the
fermion field have been evaluated, therefore $m_{f}$ and $Z_{\psi}$ do not
need any redefinition ($m_{f}$ is the physical fermionic mass and $Z_{\psi}%
=1$). The quantity $Z_{H}$ does not need to be redefined, $Z_{H}=1$, because
for each finite (no matter how large) value of the cutoff $\Lambda$ the mass
distribution of the unstable state $H$ is correctly normalized to unity. (A
redefinition of $Z_{H}$ would be necessary if the cutoff is not kept finite
but the limit $\Lambda\rightarrow\infty$ is performed first, see the
discussions and the problems in Appendix A.4.) No next-to-leading order for
the vertex is considered, therefore $g$ corresponds, in the study here
presented, to the physical value of the coupling. Finally, the bare mass
$M_{0,H}$ is chosen in such a way that (for a given cutoff $\Lambda$) the zero
of Eq. (\ref{nommass}) takes place at a fixed value of $M_{H}.$

We now turn to the spectral function of the unstable boson. We discuss first
which is the phenomenological problem we are studying: (i) the fermion mass
$m_{f}$ is known (and the quantity $2m_{f}$ sets in our model the energy
scale) (ii) the cutoff $\Lambda$ is supposed to be given and to be sizably
larger than $2m_{f};$ (iii) in a fermionic pair production process, the
measurement of the cross section, $\sigma(x)$ of Eq. (\ref{cs}), allows to
determine the function $d_{H}(x).$ In particular, one could measure the
position of the peak and its height. In this way one can fix the two
(remaining) free parameters of the model: $g$ and $M_{0,H}$. Once $g$ and
$M_{0,H}$ have been determined in order to reproduce the position and the
height of the peak of $d_{H}(x)$, the function is fixed. In principle, one can
compare the rest of its behavior with putative experimental points.

In Fig. 2 we show the spectral function for fixed values of $g=1$ and
$M_{H}=1.5$ and for two extreme values of the cutoff: $\Lambda=2$, close to
$M_{H}$, and $\Lambda=2\times10^{20}$ a situation reminiscent of the Standard
Model of particle physics where the Planck scale is much larger than any mass
of the fundamental particles. As explained before, a term in the self energy
(\ref{lupoexpl}) that mixes the cutoff energy scale and the typical energy
scale of the unstable particle is present. Thus, $d_{H}(x)$ is explicitly
dependent on the cutoff $\Lambda$: in particular, an increase of the cutoff
implies a logarithmic decrease of the width of the spectral function
$d_{H}(x).$ In turn, this means that there is not a full decoupling of the
cutoff. Notice also another remarkable property: the height of the peak do not
depend on the cutoff, which regulates solely the width of the peak.

To quantify this peculiar behavior we have done the following analysis: we
define the Breit-Wigner width $W$ of the particle as the width at half
maximum. We calculate $W$ for two values of $g$ as functions of the cutoff. In
Fig. 3 we show the corresponding results (the widths have been divided by
their values at $\Lambda=2$ to better appreciate the effect of the coupling on
the narrowing). In both cases $W$ decreases as a function of $\Lambda$, the
larger the coupling the faster is the narrowing of $W$. This result clearly
shows that it is in principle possible to determine the value of an high
energy cutoff by `measuring' the spectral function of such an unstable boson.
It is also interesting to calculate the primitives of the spectral functions
to see how the normalization is \textquotedblleft
distributed\textquotedblright\ in the energy range. We show results in Fig. 4:
for the small value $\Lambda=2$, the normalization to one is obviously reached
very close to the peak, at $x\sim3M_{H}$. On the other hand, for the large
value of the cutoff, $\Lambda=2\times10^{20}$, within an energy scale of
$\sim3M_{H}$ only roughly $50\%$ of the normalization is reached. This is
clearly due to the long high energy tail of the spectral function which is
obtained in this case. (Note that, as discussed in the Appendix A.4, by
removing the cutoff dependent terms, one obtains that most of the
normalization is distributed at extremely large energy scales, a situation
which is clearly unphysical.)

A side-remark about the time-evolution of the unstable system $\left\vert
H\right\rangle $ is in order: the survival probability amplitude $a(t)=$
$\left\langle H\left\vert e^{-iHt}\right\vert H\right\rangle $ is obtained by
computing the Fourier transform of the spectral function of the unstable state
$H$ \cite{nonexp,duecan}. The cutoff, as it has been shown in Ref.
\cite{pasca}, regulates the temporal window during which the decay law is not
exponential and possible interesting phenomena as the Quantum Zeno and
Anti-Zeno effects could arise. That is, in the present example the interval of
time, in which the survival probability $p(t)=\left\vert a(t)\right\vert ^{2}$
deviates from the exponential law, lasts for a time interval of about
$1/\Lambda.$ Thus, for a very large cutoff, the non-exponential regime elapses
for a very short time. This result is different from the corresponding one in
the case of a superrenormalizable theory, where the duration of the
non-exponential time interval is sizable and practically independent on the
cutoff \cite{nonexp}. \begin{figure}[ptb]
\begin{centering}
\epsfig{file=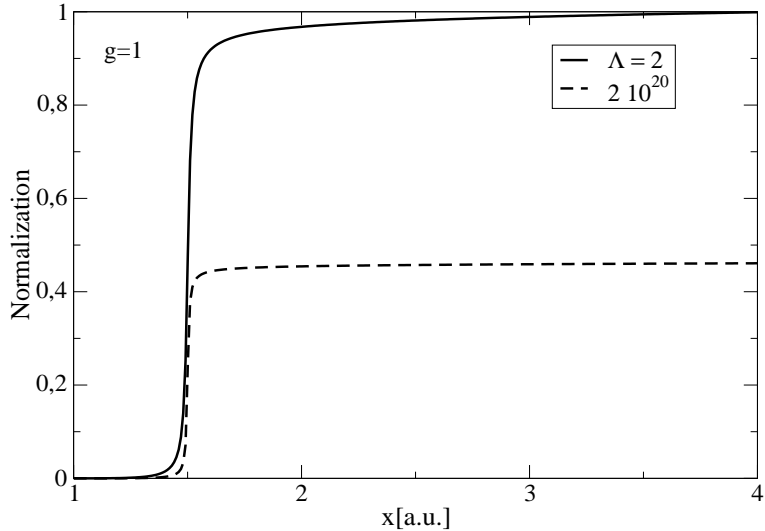,height=10cm,width=7cm,angle=-90}
\caption{Primitives of $d_H(x)$ for two values of $\Lambda$ at fixed coupling $g$. }
\end{centering}
\end{figure}

As a last result of this Section we study a numerical case which resembles the
situation of the Higgs boson. As stressed in the Introduction, for obvious
reasons our analysis cannot be considered as a realistic treatment of the
Higgs spectral function. In fact, our Lagrangian of Eq. (\ref{lint}) is by far
too simple for this purpose. Namely, while it is true that Eq. (\ref{lint}) is
one of the terms which couples the Higgs boson to a single fermion pair, many
other interaction terms are not included, such as the coupling to other
fermion pairs, loops of the Higgs field itself due to term $H^{3}$ and $H^{4}%
$, and so on. Most importantly, local gauge invariance is fully ignored in our
toy model. (For a treatment of the Higgs boson line shape in the framework of
the SM, in which dimensional regularization is used, we refer to Ref.
\cite{passarino}.) Thus, with the example of the Higgs boson, we want only to
estimate if the main result of this paper, i.e. the narrowing of the spectral
function for increasing cutoff, could be phenomenologically relevant by using,
for the the mass and the coupling constant to fermions, numerical values
compatible with the recently determined Higgs particle. For the Higgs mass we
use $M_{H}=125$ GeV \cite{lhc}. for which the Higgs couples sizably in a
$\bar{b}b$, then we use $m_{f}=m_{b}=4.18$ GeV \cite{pdg}. The coupling $g$
can be easily determined as
\[
g=\sqrt{3}\frac{m_{b}}{v}%
\]
where $v=246$ GeV is the vacuum expectation value of the Higgs field and
$\sqrt{3}$ takes into account the color degree of freedom (not present in our
Lagrangian). The tree-level decay width turns out to be $\Gamma_{H\rightarrow
\bar{\psi}\psi}^{\text{t-l}}(x=M_{H})=$ $4.28$ MeV and is thus much smaller
than the Higgs mass. The Breit-Wigner approximations is very good in this
case. For these numerical values, when varying the cutoff between the wide
range $\Lambda=$ $10^{3}$ GeV and $10^{19}$ GeV, no visible variation of the
spectral function is found. The spectral function for these numerical values
is presented in Fig. 5. (In order to obtain a visible effect one should
consider a very large and unphysical cutoff of about $10^{10000}$ GeV.)
\begin{figure}[ptb]
\begin{centering}
\epsfig{file=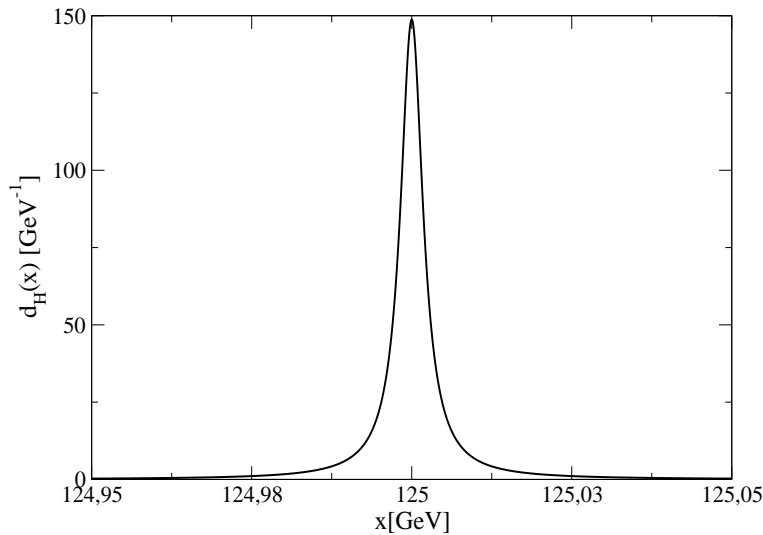,height=10cm,width=7cm,angle=-90}
\caption{Higgs spectral function obtained by considering only its decay into $b\bar{b}$. No dependence
from the cutoff is visible in this case.}
\end{centering}
\end{figure}

An hypothetical broad scalar boson, eventually also coupled to the vector
bosons, would show more visible effects: in the case of large couplings the
influence of the cutoff on the spectral function would be sizable. In some
extensions of the Standard Model, which go beyond the minimal Higgs coupling,
possible other massive scalars and broad particles are predicted, for which
the described effects could possibly be seen. Clearly, the detection of such
particles would be itself a proof of physics beyond the SM, but the effect
that we point out here is the possibility to determine the value of the cutoff
(i.e. the minimal length) by using such hypothetical broad states beyond the
SM. Presently, albeit appealing, this is only a speculative possibility.

Interestingly, the dependence on the cutoff described previously is not only a
characteristic of a scalar field coupled to fermions as in the Lagrangian of
Eq. (\ref{lag}), but would be present in each renormalizable Lagrangian. In
fact, the behavior of the decay width $\Gamma\sim x$ would be such in each
two-body decay involving a renormalizable interaction. For instance, the
coupling of the weak bosons $Z^{0}$ and $W^{\pm}$ to leptons is of such type.
Indeed, also for a detailed study of $Z^{0}$ and $W^{\pm},$ the full
complications of the SM described above should be taken into account. This is
a difficult task; an interesting intermediate step could be the study of a
vector boson coupled to fermions (and a Higgs-like particle) in the framework
of a $U(1)$ local gauge symmetric theory. While neither this study would be
realistic enough, it would constitute an attempt to take into account the
described cutoff effect by using a theoretical model which embodies some of
the most salient features of the SM. 

On the contrary, the dependence on the cutoff change if a superrenormalizable
\cite{lupo,e38} or a non-renormalizable Lagrangian \cite{lupoder} are
considered. As already mentioned, in the superrenormalizable case $\Gamma\sim
x^{-1}$ and no dependence of the spectral function on a (large) cutoff is
visible. On the other hand, for a non-renormalizable theory $\Gamma\sim x^{3}$
the theory makes sense only if the cutoff is small.

\section{Conclusions}

In this work we have computed the spectral function of a scalar boson coupled
to fermions via resummation of the one loop contributions into the scalar
propagator. The propagator satisfies the K\"{a}llen-Lehman representation and
the corresponding spectral function is normalized to unity \textit{when} a
finite (no matter how large) cutoff on the energy of the unstable boson is
used. The correct normalization is clearly connected with the completeness of
the basis of the states into which the unstable particle state is decomposed.

The finite cutoff, in turn, affects the properties of the spectral function:
the Breit-Wigner width indeed narrows as the cutoff increases. In a
fundamental theory, the existence of a energy cutoff is often connected to a
change of degrees of freedom and thus, within the Standard Model of particle
physics, the cutoff would indicate the energy scale at which `new Physics' is
expected. Another possibility is the existence of a minimal length such as the
one coming from a discrete structure of the space-time. From a
phenomenological point of view, the measurement of the line shape of an
unstable boson could signal the existence of the cutoff: phenomena occurring
at very high energy could influence low energy properties of the system. Using
the recently determined Higgs boson mass and its coupling to a fermion type
(the $b$ quark) only, we have provided a simple order of magnitude estimate of
the effect of the finite cutoff on the spectral function. It turned out that
this effect is vanishinlgy small. The situation could be different if new and
broad particles beyond SM would exist but, at the moment, this is only a
speculative possibility.

Interestingly, also hadronic physics, where the cutoff is related to the scale
of confinement, some interesting modification of the spectral function could
have a phenomenological relevance, for instance for the medium modification of
the $\rho$ meson spectral function which heavy ions experiments are looking for.

There are two possible future directions to be taken: on the theoretical side,
one should go beyond the resummed fermionic one-loop approximation considered
here. The next-to-leading order correction for the propagator of the unstable
state $H$ arises when considering an insertion of a scalar virtual exchange in
the fermionic loop. This diagram, which is of order $g^{4}$, should be also
resummed. Being the scalar propagator also part of the dressing, one has a
problem of the Dyson-Schwinger type, which is obviously more difficult to
solve and the renormalization of the charge would be necessary within this
context. Moreover, also the renormalization of the charge would be necessary
at this order. Anyhow, considering that the presented considerations about the
finiteness of the cutoff are rather general, we do not expect a qualitative
change of our results. However, a mathematical proof that this is the case
would be very valuable. On the phenomenological side, we plan to compute the
spectral functions of vector bosons coupled to fermions, which is potentially
relevant for the weak interaction. Namely, the presented phenomena are quite
general for each renormalizable theory, and therefore should apply to the weak
gauge bosons. In this way we can study the effect of a putative cutoff in the
weak sector as well.

In the end, we think that it would be also possible to check our results using
lattice Quantum Field Theory. In that case, a cutoff is naturally present (the
finite lattice spacing) which resembles closely our finite cutoff used here.
The simulation of the simple Yukawa Lagrangian in Eq. (\ref{lint}) should be
feasible also in Minkowski space.

\bigskip

\textbf{Acknowledgments: } G.P. acknowledges financial support from the
Italian Ministry of Research through the program \textquotedblleft Rita Levi
Montalcini\textquotedblright. F. G. thanks the Foundation of the Polytechnical
Society of Frankfurt for support through an Educator fellowship.

\appendix

\section{Details of the calculations}

\subsection{The decay width}

The decay width is explicitly evaluated by making use of the following
standard relations:%
\begin{align}
\Lambda_{+}(\vec{k})_{ab}  &  =\sum_{\alpha}u_{a}^{(\alpha)}(\vec{k}%
)\overline{u}_{b}^{(\alpha)}(\vec{k})=\left(  \frac{\gamma^{\mu}k_{\mu}+m_{f}%
}{2m_{f}}\right)  _{ab}\text{ ,}\\
\Lambda_{\_}(\vec{k})_{ab}  &  =-\sum_{\alpha}v_{a}^{(\alpha)}(\vec{k})\bar
{v}_{b}^{(\alpha)}(\vec{k})=\left(  \frac{-\gamma^{\mu}k_{\mu}+m_{f}}{2m_{f}%
}\right)  _{ab}\text{ .}%
\end{align}
Out of the latter expressions, one rewrites the squared amplitude as:%
\begin{align}
\sum_{\alpha,\beta}\left\vert \mathcal{M}^{\alpha\beta}\right\vert ^{2}  &
=\sum_{\alpha,\beta}g^{2}\left[  \bar{u}_{a}^{(\alpha)}(\vec{k}_{1}%
)v_{a}^{(\beta)}(\vec{k}_{2})\right]  \left[  \bar{u}_{b}^{(\alpha)}(\vec
{k}_{1})v_{b}^{(\beta)}(\vec{k}_{2})\right]  ^{\dagger}\\
&  =g^{2}\sum_{\alpha,\beta}\left[  \bar{u}_{a}^{(\alpha)}(\vec{k}_{1}%
)v_{a}^{(\beta)}(\vec{k}_{2})\right]  \left[  \bar{v}_{b}^{(\beta)}(\vec
{k}_{2})u_{b}^{(\alpha)}(\vec{k}_{1})\right] \\
&  =-g^{2}\Lambda_{+}(\vec{k}_{1})_{ba}\Lambda_{\_}(\vec{k}_{2})_{ab}%
=g^{2}Tr\left[  \frac{\gamma^{\mu}k_{1\mu}+m_{f}}{2m_{f}}\frac{\gamma^{\nu
}k_{2\nu}-m_{f}}{2m_{f}}\right] \\
&  =g^{2}\frac{4(k_{1}\cdot k_{2})-4m_{f}^{2}}{4m_{f}^{2}}=g^{2}\frac{8\left(
\frac{x^{2}}{4}-m_{f}^{2}\right)  }{4m_{f}^{2}}\text{ .}%
\end{align}
where in the last step we have taken into account that%
\begin{equation}
k_{1}\cdot k_{2}=\frac{x^{2}-2m_{f}^{2}}{2}\text{ .}%
\end{equation}

\subsection{Regularizations}

\subsubsection{Cutoff scheme}

Here we show which is the formal expression of the nonlocal Lagrangian
necessary to generate the cutoff vertex function described in Sec. 2:%

\begin{equation}
\left(  \mathcal{L}_{int}\right)  _{cutoff}=g\int d^{4}zd^{4}y_{1}d^{4}%
y_{1}H(x+z)\bar{\psi}(x+y_{1})\psi(x+y_{2})\tilde{\phi}(z,y_{1},y_{2})\text{
,}%
\end{equation}
where the vertex-function $\tilde{\phi}(z,y_{1},y_{2})$ in position space has
been introduced. The case $\tilde{\phi}(z,y_{1},y_{2})$ $=\delta
(z)\delta(y_{1})\delta(y_{2})$ delivers the local limit of the Lagrangian
(\ref{lag}). (For similar approaches see Refs. \cite{lyubo,nonlocal} and refs. therein.)

By performing the usual steps, we obtain that the vertex function in momentum
space is given by the Fourier transform of $\tilde{\phi}(z,y_{1},y_{2})$:%
\begin{equation}
\phi(p,k_{1},k_{2})=\int d^{4}zd^{4}y_{1}d^{4}y_{2}e^{ipz}e^{-ik_{1}y_{1}%
}e^{-ik_{2}y_{2}}\tilde{\phi}(z,y_{1},y_{2})\text{ .}%
\end{equation}
Here we assume that $\tilde{\phi}(z,y_{1},y_{2})$ is such that
\begin{equation}
\phi(p,k_{1},k_{2})=\phi_{\Lambda}\left(  p,q=\frac{k_{1}-k_{2}}{2}\right)
=f_{\Lambda}\left(  \frac{q^{2}p^{2}-(q\cdot p)^{2}}{p^{2}}\right)  \text{ .}%
\end{equation}
Note, to this end $\tilde{\phi}(z,y_{1},y_{2})$ must be of the form
$\tilde{\phi}(z,y_{1},y_{2})=\varphi(z,y_{1}-y_{2})\delta(y_{1}+y_{2}).$ In
fact, introducing $y=y_{1}-y_{2}$ and $Y=y_{1}+y_{2},$ one finds
\begin{align}
\phi(p,k_{1},k_{2})  &  =\int d^{4}zd^{4}yd^{4}Ye^{ipz}e^{-ik_{1}y_{1}%
}e^{-ik_{2}y_{2}}\varphi(z,y)\delta(Y)\nonumber\\
&  =\text{ }\int d^{4}zd^{4}yd^{4}Ye^{ipz}e^{-i(k_{1}-k_{2})y}e^{-i(k_{1}%
+k_{2})Y}\varphi(z,y)\delta(Y)\nonumber\\
&  =\int d^{4}zd^{4}ye^{ipz}e^{-i2qy}\varphi(z,y)=\phi_{\Lambda}\left(
p,q\right)  \text{ .}%
\end{align}

Moreover, In\ Sec. 2 we worked with a bare mass $M_{0,H}$ and a `physical'
mass $M_{H}$. Alternatively, one could work with the inclusion of counterterms
and impose that the quantity $M_{H}$ entering in the Lagrangian is the nominal
mass of the resonance (thus $M_{0,H}=M_{H}$). In the present case one
introduces the counterterm
\begin{equation}
\mathcal{L}_{ct}=-\frac{g^{2}\operatorname{Re}\Sigma(M_{0,H})}{2}H^{2}\text{
.}%
\end{equation}
Considering at the one-loop level the Lagrangian%
\begin{equation}
\mathcal{L}_{cutoff,counterterms}\mathcal{=L}_{0}\mathcal{+}\left(
\mathcal{L}_{int}\right)  _{cutoff}+\mathcal{L}_{ct}\text{ ,}%
\end{equation}
where $\mathcal{L}_{0}$ describes the free Lagrangian, Eq. (\ref{nommass})
takes the modified form
\begin{equation}
x^{2}-M_{0,H}^{2}+g^{2}\operatorname{Re}\Sigma(x^{2})-g^{2}\operatorname{Re}%
\Sigma(M_{0,H})=0\text{ ,}%
\end{equation}
thus implying the solution
\begin{equation}
x=M_{H}=M_{0,H}\text{ .}%
\end{equation}
Obviously, nothing substantial would change by following this procedure. Note,
the introduction of counterterms can be applied to other regularization
schemes as well.

\subsubsection{Pauli-Villars}

In the Pauli-Villars (PV) approach one subtracts from the original loop of
particles with mass $m_{f}$ a second loop with particles of mass $\Lambda
_{PV}\gg m_{f}$:%

\begin{equation}
\Sigma(x)\rightarrow\Sigma_{PV}(x)=\lim_{\Lambda\rightarrow\infty}\left(
\Sigma(x)-\left(  \Sigma(x)\right)  _{m_{f}\rightarrow\Lambda_{PV}}\right)
\text{ .}%
\end{equation}
Formally, we can still use the previous expression with the cutoff $\Lambda$,
but here the condition $m_{f}\ll\Lambda_{PV}\ll\Lambda$ must hold. In the end
each dependence on $\Lambda$ disappears and its value can be sent to infinity,
but the dependence on the new high scale $\Lambda_{PV}$ is present. The
explicit expression for $\Sigma_{PV}(x)$ reads:
\begin{align}
\Sigma_{PV}(x)  &  =-\frac{\left(  x^{2}-4m_{f}^{2}\right)  ^{3/2}}{4\pi^{2}%
x}\text{arctanh}\left(  \frac{x}{\sqrt{x^{2}-4m_{f}^{2}}}\right)  +\frac
{x^{2}}{4\pi^{2}}\ln\left(  \frac{\Lambda_{PV}}{m_{f}}\right) \nonumber\\
&  +\frac{\left(  x^{2}-4\Lambda_{PV}^{2}\right)  ^{3/2}}{4\pi^{2}%
x}\text{arctanh}\left(  \frac{x}{\sqrt{x^{2}-4\Lambda_{PV}^{2}}}\right)
+\text{const .}%
\end{align}
As long as $x\ll\Lambda_{PV},$ this result is equivalent to the form
(\ref{lupoappr}), also including the term proportional to $x^{2}\ln
\Lambda_{PV}.$ Thus, if one numerically sets $\Lambda=\Lambda_{PV}\gg m_{f}$
one finds, in the vicinity of the peak, a behavior which is very similar to
the one obtained in the cutoff case: also the narrowing of the spectral
function is obtained. However, for values of $x$ comparable to $\Lambda_{PV}$,
the loop contribution $\Sigma_{PV}(x)$ is modified due to the fact that the
additional degree of freedom related to the `particle' with mass $\Lambda
_{PV}$ becomes active.

Moreover, the normalization of the spectral function to unity is not
fulfilled. We can easily understand what goes wrong in the present case by
writing the modified Lagrangian which delivers the Pauli-Villars formulae:%

\begin{align}
\left(  \mathcal{L}_{H\psi}\right)  _{PV}  &  =\frac{1}{2}(\partial_{\mu
}H)^{2}-\frac{1}{2}M_{0,H}^{2}H^{2}+\bar{\psi}(i\gamma^{\mu}\partial_{\mu
}-m_{f})\psi+\bar{\psi}_{PV}(i\gamma^{\mu}\partial_{\mu}-\Lambda_{PV}%
)\psi_{PV}\nonumber\\
&  +gH\bar{\psi}\psi+igH\bar{\psi}_{PV}\psi_{PV}\text{ .}%
\end{align}
The new fermion field $\psi_{PV}$ with mass $\Lambda_{PV}$ is introduced: it
should be noticed that, in order to obtain the required cancellation, the
coupling of the latter with the boson field $H$ is an imaginary number $ig.$
For this reason, the $S$ matrix is not unitary. As a consequence, the
normalization of $d_{H}(x)$ is lost (the `new' particle gives rise to a
negative contribution to the spectral function). In conclusion, the use of the
Pauli-Villars scheme delivers similar results to the cutoff scheme as long as
$\Lambda_{PV}$ is finite, but we prefer the latter because it explicitly
guarantees the correct normalization of the spectral function to unity.
Conversely, sending $\Lambda_{PV}$ to infinity generates the same problems
discussed in Secs. 2.4 and A.4.

\subsubsection{Dimensional regularization}

Within the dimensional regularization scheme, one calculates the integral of
the fermion loop in $d$ dimensions with $d=4-\epsilon$ and then takes the
limit $\epsilon\rightarrow0$. In this case the coupling constant $g$ has the
dimension of [Energy$^{\epsilon}$], therefore the spectral function should
scale as $x^{-1-2\epsilon},$ which is convergent for each -no matter how
small- value of $\epsilon.$ This is mathematically reminiscent to the finite
cutoff case, but the physical interpretation of a nonzero $\epsilon$ is not meaningful.

When calculating the self energy $\Sigma(x)$ using the standard formulae (see
also Eq. 10.33 of \cite{peskin}) one finds:%

\begin{equation}
\Sigma(x)\rightarrow\Sigma_{DR}(x)=4i\int_{0}^{1}dy\int\frac{d^{d}l}%
{(2\pi)^{d}}\frac{\Delta+l^{2}}{(l^{2}-\Delta)^{2}}\text{ ;}%
\end{equation}
then, after integrating in $d^{d}l$ we obtain:
\begin{equation}
\Sigma(x)_{DR}=-4\int_{0}^{1}dy\frac{1}{(4\pi)^{d/2}}\left(  \Delta
\Gamma(2-d/2)\frac{1}{\Delta^{2-d/2}}-2\Gamma(1-d/2)\frac{1}{\Delta^{1-d/2}%
}\right)  \text{ ,}%
\end{equation}
where $\Delta=m_{f}^{2}-y(1-y)x^{2}$ and where $\Gamma$ is the Euler function.
Making use of Eqs. A49 and A50 of \cite{peskin} we obtain:%

\begin{equation}
\Sigma(x)_{DR}=-12\int_{0}^{1}dy\frac{1}{(4\pi)^{2}}\Delta(2/\epsilon
-\gamma+O(\epsilon))\left(  1-\frac{\epsilon}{2}\log\Delta\right)
\end{equation}
Calculating the integral and keeping the leading terms in $\epsilon$ for the
real part we get:%

\begin{equation}
\Sigma(x)_{DR}=\frac{x^{2}-6m_{f}^{2}}{4\pi^{2}\epsilon}-\frac{\left(
x^{2}-4m_{f}^{2}\right)  ^{3/2}}{4\pi^{2}x}\text{arctanh}\left(  \frac
{x}{\sqrt{x^{2}-4m_{f}^{2}}}\right)  \label{lupodr}%
\end{equation}
Notice that the imaginary part obtained in this scheme is, as it should, equal
to the one obtained in the other schemes when the cutoff is sent to infinity.
Comparing the previous equation with Eq. (\ref{lupoappr}) one sees the
correspondence $1/\epsilon\propto\ln\Lambda.$ The divergence of the real part
is as well known linear and non logarithmic and it is usually reabsorbed in
the mass and field strength renormalization in the case of stable particles.
In the case of unstable particles, in the so called \textquotedblleft complex
mass renormalization scheme\textquotedblright\ \cite{denner}, the
renormalization procedure is performed by introducing a complex mass for the
resonance. For the Higgs particle for instance, to simplify the calculation,
one expands the mass counterterm for $\Gamma_{H}/M_{H}\rightarrow0$, an
approximation which definitely holds being the mass of the Higgs only $125$
GeV. However, eliminating the term proportional to $1/\epsilon$ is completely
equivalent to neglect the term proportional to $x^{2}\log\Lambda$ in Eq.
(\ref{lupoappr}). It generates many inconsistencies, as mentioned in Sec. 2.4
and shown in Sec. A.4.

\subsection{Correct normalization to unity of the spectral function in
presence of a cutoff}

Let us consider the state $\left\vert H\right\rangle $ as the eigenstate of
the unperturbed Hamiltonian $H_{0}$ which fulfills the normalization condition
$\left\langle H|H\right\rangle =1.$ The full set of eigenstates of the
Hamiltonian $H$ reads $\{\left\vert x\right\rangle \}$ with $H\left\vert
x\right\rangle =x\left\vert x\right\rangle $ and $x\geq0$. Expressing
$\left\vert H\right\rangle $ in terms of $\left\vert x\right\rangle $ implies%
\begin{equation}
\left\vert H\right\rangle =\int_{0}^{\infty}\mathrm{dx}a(x)\left\vert
x\right\rangle \text{ .}%
\end{equation}
The quantity $d_{H}(x)=\left\vert a(x)\right\vert ^{2}$ is the `spectral
function' which is evaluated in this work as the imaginary part of the
propagator (see also Ref. \cite{duecan} for a more detailed discussion of
these relations). It naturally follows that%
\begin{equation}
1=\left\langle H|H\right\rangle =\int_{0}^{\infty}\mathrm{dx}d_{H}(x)\text{ .}%
\end{equation}

Taking into account that, in the case of a hard cutoff, $d_{H}(y)$ vanishes
for $y>2\sqrt{\Lambda^{2}+m_{f}^{2}}$, the K\"{a}llen-Lehman representation
can be rewritten as%
\begin{equation}
\Delta_{H}(x)=\int_{0}^{2\sqrt{\Lambda^{2}+m_{f}^{2}}}\mathrm{dy}\frac
{d_{H}(y)}{x^{2}-y^{2}+i\varepsilon}\text{ .}%
\end{equation}
For $x\gg2\sqrt{\Lambda^{2}+m_{f}^{2}}$ no pole is encountered in the
integral; at the same time the loop function $\Sigma(x)$ is very small (the
real part goes to zero very fast for $x\gg2\sqrt{\Lambda^{2}+m_{f}^{2}}$ while
the imaginary part is identically zero). It then follows
\begin{equation}
\frac{1}{x^{2}}=\frac{1}{x^{2}}\int_{0}^{2\sqrt{\Lambda^{2}+m_{f}^{2}}%
}\mathrm{dy}d_{H}(y)\text{ for }x\gg2\sqrt{\Lambda^{2}+m_{f}^{2}}\text{ ,}%
\end{equation}
that is:%
\begin{equation}
\int_{0}^{2\sqrt{\Lambda^{2}+m_{f}^{2}}}\mathrm{dy}d_{H}(y)=\int_{0}^{\infty
}\mathrm{dy}d_{H}(y)=1\text{ .}%
\end{equation}
We now turn to the case of a smooth cutoff function, which assures that the
loop contribution $\Sigma(x)$ is very small beyond a certain energy scale
$\Lambda.$ However, the imaginary part, and so the spectral function, do not
vanish exactly for $x\gg\Lambda$. The proof of the correct normalization is in
this case more difficult. As a first step, we decompose the integral as
\begin{equation}
\Delta_{H}(x)=\int_{0}^{m_{f}\sqrt{x}}\mathrm{dy}\frac{d_{H}(y)}{x^{2}%
-y^{2}+i\varepsilon}+\int_{m_{f}\sqrt{x}}^{\infty}\mathrm{dy}\frac{d_{H}%
(y)}{x^{2}-y^{2}+i\varepsilon}\text{ .}%
\end{equation}
When $m_{f}\sqrt{x}\gg\Lambda$ (which implies also $x\gg\Lambda$) the
propagator is $\Delta_{H}(x)=1/x^{2}.$ We thus obtain:%
\begin{equation}
\frac{1}{x^{2}}=\frac{1}{x^{2}}\int_{0}^{m_{f}\sqrt{x}}\mathrm{dy}%
d_{H}(y)+\text{P}\int_{m_{f}\sqrt{x}}^{\infty}\mathrm{dy}\frac{d_{H}(y)}%
{x^{2}-y^{2}+i\varepsilon}\text{ }%
\end{equation}
where P stands for principal part. In the large energy limit the spectral
function $d_{H}(y)$ can be approximated as
\begin{equation}
d_{H}(y)\simeq\frac{g^{2}}{8\pi y}f_{\Lambda}^{2}\left(  -\sqrt{\frac{y^{2}%
}{4}-m_{f}^{2}}\right)  \simeq\frac{g^{2}}{8\pi y}f_{\Lambda}^{2}\left(
-y/2\right)  \text{ .}%
\end{equation}
Finally, taking the limit $x\rightarrow\infty$ we obtain%
\begin{equation}
1=\int_{0}^{\infty}\mathrm{dy}d_{H}(y)+\lim_{x\rightarrow\infty}\left[
x^{2}\text{P}\int_{m_{f}\sqrt{x}}^{\infty}\mathrm{dy}\frac{g^{2}}{8\pi y}%
\frac{f_{\Lambda}^{2}\left(  -y/2\right)  }{x^{2}-y^{2}+i\varepsilon}\right]
\text{ .}%
\end{equation}
If the cutoff function is such that
\begin{equation}
\lim_{x\rightarrow\infty}\left[  x^{2}\text{P}\int_{m_{f}\sqrt{x}}^{\infty
}\mathrm{dy}\frac{g^{2}}{8\pi y}\frac{f_{\Lambda}^{2}\left(  -y/2\right)
}{x^{2}-y^{2}+i\varepsilon}\right]  =0 \label{pp}%
\end{equation}
it follows that the correct normalization condition holds:
\begin{equation}
1=\int_{0}^{\infty}\mathrm{dy}d_{H}(y)\text{ .}%
\end{equation}
Indeed, as long as $f_{\Lambda}^{2}\left(  -y/2\right)  $ falls off
sufficiently fast, Eq. (\ref{pp}) is fulfilled. A power-like or exponential
decrease introduced to assure the convergence of the loop integral
automatically implies the validity of Eq. (\ref{pp}), hence the correct
normalization to unity of the spectral function (independently, also here, on
the precise value of the cutoff).

\subsection{Removing completely the $\Lambda$ dependence}

We describe here the standard renormalization of the Lagrangian under study in
the case of unstable particles. The starting point is the loop expression in
Eq. (\ref{lupoappr}). First, we rewrite the real and the imaginary part of
$\tilde{\Sigma}(x)$ as follows:
\begin{align}
\operatorname{Re}[\tilde{\Sigma}(x)]  &  =A(x)+\frac{\Lambda^{2}}{2\pi^{2}%
}-\frac{6m_{f}^{2}}{4\pi^{2}}\ln\left(  \frac{2\Lambda}{m_{f}}\right)
+\frac{x^{2}}{4\pi^{2}}\ln\left(  \frac{2\Lambda}{m_{f}}\right) \\
\left\vert \operatorname{Im}[\tilde{\Sigma}(x)]\right\vert  &  =\frac{x}%
{g^{2}}\Gamma_{H\rightarrow\bar{\psi}\psi}^{\text{t-l}}(x)=\text{ }%
\frac{\left(  \frac{x^{2}}{4}-m_{f}^{2}\right)  ^{3/2}}{\pi x}\theta
(x-2m_{f})\text{ ,}%
\end{align}
where the cutoff independent quantity $A(x)$ is given by
\begin{equation}
A(x)=-\frac{\left(  x^{2}-4m_{f}^{2}\right)  ^{3/2}}{8\pi^{2}x}\ln\left(
\frac{x+\sqrt{x^{2}-4m_{f}^{2}}}{x-\sqrt{x^{2}-4m_{f}^{2}}}\right)  \text{ .}%
\end{equation}
The propagator takes the form%

\begin{align}
\tilde{\Delta}_{H}(x)  &  =\left[  x^{2}-M_{0,H}^{2}+g^{2}A(x)+\frac
{g^{2}\Lambda^{2}}{2\pi^{2}}\right. \nonumber\\
&  \left.  -\frac{6g^{2}m_{f}^{2}}{4\pi^{2}}\ln\left(  \frac{2\Lambda}{m_{f}%
}\right)  +g^{2}\frac{x^{2}}{4\pi^{2}}\ln\left(  \frac{2\Lambda}{m_{f}%
}\right)  +ig^{2}\operatorname{Im}[\tilde{\Sigma}(x)]+i\varepsilon\right]
^{-1}\text{ .}%
\end{align}
The renormalized mass $M_{H,ren}$ is defined as the solution of the equation
in a similar way as Eq. (\ref{nommass}):
\begin{equation}
M_{H,ren}^{2}-M_{0,H}^{2}+g^{2}A(M_{H,ren})+\frac{g^{2}\Lambda^{2}}{2\pi^{2}%
}-\frac{6g^{2}m_{f}^{2}}{4\pi^{2}}\ln\left(  \frac{2\Lambda}{m_{f}}\right)
+g^{2}\frac{M_{H,ren}^{2}}{4\pi^{2}}\ln\left(  \frac{2\Lambda}{m_{f}}\right)
=0\text{ .}%
\end{equation}
By performing a Taylor expansion of the real part around $M_{H,ren}$ we obtain%
\begin{align}
\tilde{\Delta}_{H}(x)  &  =\left[  \left(  x^{2}-M_{H,ren}^{2}\right)  \left(
1+g^{2}\left(  \partial_{x^{2}}A\right)  _{x^{2}=M_{H,ren}^{2}}+\frac{g^{2}%
}{4\pi^{2}}\ln\left(  \frac{2\Lambda}{m_{f}}\right)  \right)  \right.
\nonumber\\
&  \left.  +g^{2}\tilde{A}(x)+ig^{2}\operatorname{Im}[\tilde{\Sigma
}(x)]+i\varepsilon\text{ }\right]  ^{-1}%
\end{align}
where
\begin{equation}
\tilde{A}(x)=A(x)-A(x=M_{H,ren})-\left(  \partial_{x^{2}}A\right)
_{x^{2}=M_{H,ren}^{2}}\left(  x^{2}-M_{H,ren}^{2}\right)  \text{ .}%
\end{equation}

By introducing the wave function renormalization $H\rightarrow\sqrt{Z_{H}}H$
the propagator takes the form%
\begin{align}
\tilde{\Delta}_{H}(x)  &  =\frac{1}{Z_{H}}\left[  \left(  x^{2}-M_{H,ren}%
^{2}\right)  \left(  1+g^{2}\left(  \partial_{x^{2}}A\right)  _{x^{2}%
=M_{H,ren}^{2}}+\frac{g^{2}}{4\pi^{2}}\ln\left(  \frac{2\Lambda}{m_{f}%
}\right)  \right)  \right. \nonumber\\
&  \left.  +g^{2}\tilde{A}(x)+ig^{2}\operatorname{Im}[\tilde{\Sigma
}(x)]+i\varepsilon\right]  ^{-1}\\
&  =\frac{1}{Z_{H}K}\left[  \left(  x^{2}-M_{H,ren}^{2}\right)  +\frac{g^{2}%
}{K}\tilde{A}(x)+i\frac{g^{2}}{K}\operatorname{Im}[\tilde{\Sigma
}(x)]+i\varepsilon\right]  ^{-1}\text{ .}%
\end{align}
where the (formally divergent) quantity $K$ reads%
\begin{equation}
K=1+g^{2}\left(  \partial_{x^{2}}A\right)  _{x^{2}=M_{H,ren}^{2}}+\frac{g^{2}%
}{4\pi^{2}}\ln\left(  \frac{2\Lambda}{m_{f}}\right)  \text{ .}%
\end{equation}
Now, one needs also to perform a renormalization of the coupling $g\rightarrow
g_{ren}$ (which is obtained here through multiplicative constant, thus no
running coupling arises at this level), leading to the following equations for
$Z_{H}$ and $g_{ren}$:
\begin{align}
Z_{H}K  &  =N\text{ ,}\\
\frac{g^{2}}{K}  &  =g_{ren}^{2}\text{ ,}%
\end{align}
whereas $N$ and $g_{ren}$ are finite constant. In this way the propagator
takes the renormalized form%
\begin{equation}
\tilde{\Delta}_{H}(x)=\frac{1}{N}\left[  \left(  x^{2}-M_{H,ren}^{2}\right)
+g_{ren}^{2}\tilde{A}(x)+ig_{ren}^{2}\operatorname{Im}[\tilde{\Sigma
}(x)]+i\varepsilon\right]  ^{-1}\text{ }%
\end{equation}
in which the dependence on $\Lambda$ has been completely eliminated. One might
think that each problem is solved here, but this is not the case. Tho show it
we turn our attention to the spectral function:
\begin{equation}
\tilde{d}_{H}(x)=\lim_{\varepsilon\rightarrow0^{+}}\frac{2x}{\pi}\left\vert
\operatorname{Im}\tilde{\Delta}_{H}(x)\right\vert \text{ .}%
\end{equation}
For $x\gg2m_{f}$ (that is, away from threshold) the following simplifications
are valid:
\begin{align}
\tilde{A}(x)  &  =-\frac{x^{2}}{4\pi^{2}}\ln x\text{ ,}\\
x\Gamma_{H\rightarrow\bar{\psi}\psi}^{\text{t-l,ren}}(x)  &  =\text{ }%
\frac{x^{2}}{8\pi}\text{.}%
\end{align}
Then, the spectral function for $x\gg2m_{f}$ is approximated by the following
expression:%
\begin{equation}
\tilde{d}_{H}(x)\simeq\text{ }\frac{1}{N}\frac{g_{ren}^{2}x^{3}}{8\pi}\frac
{1}{\left(  x^{2}-\frac{g_{ren}^{2}x^{2}}{4\pi^{2}}\ln x\right)  ^{2}+\left(
\frac{x^{2}g_{ren}^{2}}{8\pi}\right)  ^{2}}\text{ .}%
\end{equation}
Let us define the point $x_{\ast}$ as the solution of the following
transcendental equation:
\begin{equation}
1=\frac{g_{ren}^{2}}{4\pi^{2}}\ln x_{\ast}\text{ }\rightarrow x_{\ast
}=e^{\frac{4\pi^{2}}{g_{ren}^{2}}}\text{ .}%
\end{equation}
Then, for $2m_{f}\ll x\lesssim x_{\ast}$ the function $\tilde{d}_{H}(x)$
scales as $1/x$ and for $x\gtrsim x_{\ast}$ the logarithm starts to dominate
and $\tilde{d}_{H}(x)$ scales as $1/(x\ln^{2}x),$ which assures a (slow)
convergence of the integral $\int_{0}^{\infty}d_{H}(x)dx.$ A numerical
evaluation shows that the following approximately scaling law holds:%

\begin{equation}
\int_{0}^{\infty}\tilde{d}_{H}(x)dx\simeq\frac{80}{Ng_{ren}^{2}}\text{ .}%
\end{equation}

At a general level we can immediately discuss two basic problems of the
spectral function $\tilde{d}_{H}(x)$:

(a) The \textquotedblleft distribution\textquotedblright\ of the normalization
in the energy range of the particle is spread over very large value of $x$. In
Fig. 6, we show the quantity $F(x)=N\int_{100}^{x}\tilde{d}_{H}(y)dy$ for two
values of the coupling constant $g$ (the value $100$ as the lower limit of
integration corresponds to a value much larger than the peak position and is
suited to study the energy localization of the state faraway from the peak).
The saturation of $F(x)$ is reached only at extremely high energy, very far
from the nominal mass of the particle. This fact is clearly connected to the
slow logarithmic convergence of the integral of the spectral function.

(b) The effect of changing coupling is evident: the smaller the coupling
$g_{ren}$, the larger the normalization is and the later is reached. This
property also implies that the small $g_{ren}$ limit is completely at odds
with the basic expectation of having the unstable particle mostly localized
around the peak.

\begin{figure}[ptb]
\begin{centering}
\epsfig{file=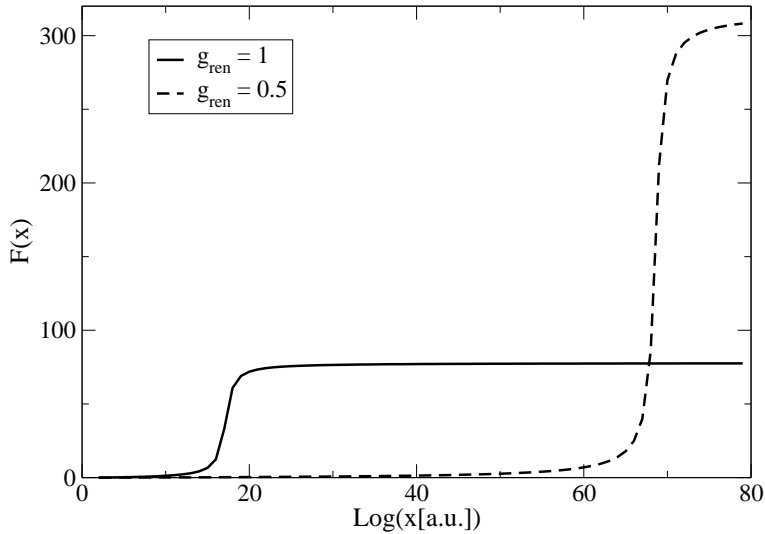,height=10cm,width=7cm,angle=-90}
\caption{The integral function $F(x)$ is displayed for two values of the coupling. The normalization of the
spectral function (not normalized to one) is obtained at extremely large energies.}
\end{centering}
\end{figure}

In order to evaluate the spectral function one has to determine the
normalization condition $N.$ To this end, it is useful to recall the case of a
stable scalar state: $M_{H,ren}<2m_{f}$, which implies $\operatorname{Im}%
[\tilde{\Sigma}(x=M_{H,ren})]=0.$ In this case, the requirement is that the
free propagator with residue $1$ at the pole is obtained, thus: $N=1,$ which
is a clear and physically meaningful requirement: The state $H$ is a stable
asymptotic state entering, for instance, as initial or final state in a
two-body process, for which the canonical normalization holds. However, in our
case we deal with an unstable state for which $M_{H,ren}>2m_{f}$: there is no
pole below threshold and it is not clear which condition should be used. Two
possibilities are the following:

(i) $N=1.$ In this way the coefficient multiplying the term $\left(
x^{2}-M_{H,ren}^{2}\right)  $ in the denominator of the propagator is unity;
this represents a simple generalization of the stable case. However, setting
$N=1$ means that the normalization to unity of $\int_{0}^{\infty}\tilde{d}%
_{H}(x)dx$ is in general lost and one violates a very basic property of
Quantum Mechanics. We regard this `solution' as unphysical. It would rather
corresponds to an \emph{ad hoc} prescription to ignore the problems. Notice
that within this prescription the curve in the vicinity of the peak looks very
similar to the case of a not too large but finite cutoff.

(ii) One can set $N$ as being dependent on $g$, in such a way that $\int
_{0}^{\infty}\tilde{d}_{H}(x)dx=1$ is fulfilled. Still, as discussed above,
the amount of the integral in the vicinity of the peak represents only a very
small contribution to the normalization of the spectral function (due to the
slow convergence of the latter). Additionally, if we aim to describe the
situation in which the spectral function has a certain given (putative
measured) height for $x=M_{H,ren}=1.5$, one runs into problems because the
quantity $\tilde{d}_{H}(x=M_{H,ren})$ is practically independent on $g_{ren}$
which is a quite unrealistic feature.

In conclusion, we believe that the here outlined procedure is not physical.


\begin{thebibliography}{99}                                                                                               %


\bibitem {salam}P.~T.~Matthews and A.~Salam,
Phys.\ Rev.\ \textbf{112} (1958) 283.
P.~T.~Matthews and A.~Salam,
Phys.\ Rev.\ \textbf{115} (1959) 1079.


\bibitem {Achasov:2004uq}N.~N.~Achasov and A.~V.~Kiselev,
Phys.\ Rev.\ D \textbf{70} (2004) 111901 [hep-ph/0405128].


\bibitem {lupo}F.~Giacosa, G.~Pagliara,
Phys.\ Rev.\ \textbf{C76 } (2007) 065204. [arXiv:0707.3594 [hep-ph]].

\bibitem {peskin}Peskin, M.~E. and Schroeder, D.~V. (1995). \emph{An
Introduction to Quantum Field Theory} (Addison-Wesley, Oxford).

\bibitem {weinberg}Weinberg, S. (1996). \emph{The Quantum Theory of Fields}
(Cambridge University Press).

\bibitem {itzy}Itzykson, C. and Zuber, J.~B. (1980). \emph{Quantum field
theory} (McGraw-Hill, New York).

\bibitem {zee}A.~Zee, \newblock {\em Quantum Field Theory in a Nutshell}
(Princeton University Press, Princeton, NJ, 2003).

\bibitem {terning}J.~Terning,
Phys.\ Rev.\ D \textbf{44} (1991) 887.

\bibitem {lyubo}
A.~Faessler, T.~Gutsche, M.~A.~Ivanov, V.~E.~Lyubovitskij and P.~Wang,
Phys.\ Rev.\ D \textbf{68} (2003) 014011 [arXiv:hep-ph/0304031].


\bibitem {djouadi}A.~Djouadi,
Phys.\ Rept.\ \textbf{457} (2008) 1 [hep-ph/0503172].


\bibitem {chpt}J.~Gasser and H.~Leutwyler,
Annals Phys.\ \textbf{158}, 142 (1984);


\bibitem {pich}A.~Pich,
Rept.\ Prog.\ Phys.\ \textbf{58} (1995) 563 [hep-ph/9502366].

\bibitem {klevansky}S.~P.~Klevansky,
Rev.\ Mod.\ Phys.\ \textbf{64} (1992) 649.


\bibitem {oldsm}M.~Gell-Mann and M.~Levy,
Nuovo Cim.\ \textbf{16}, 705 (1960);


\bibitem {dick}
D.~Parganlija, P.~Kovacs, G.~Wolf, F.~Giacosa and D.~H.~Rischke,
arXiv:1208.0585 [hep-ph].


\bibitem {passarino}S.~Goria, G.~Passarino and D.~Rosco,
arXiv:1112.5517 [hep-ph].



\bibitem {denner}A.~Denner, S.~Dittmaier, M.~Roth and L.~H.~Wieders,
Nucl.\ Phys.\ B \textbf{724} (2005) 247 [Erratum-ibid.\ B \textbf{854} (2012)
504] [hep-ph/0505042].




\bibitem {denner2}A.~Denner and S.~Dittmaier,
Nucl.\ Phys.\ Proc.\ Suppl.\ \textbf{160} (2006) 22 [hep-ph/0605312].




\bibitem {denner3}T.~Bauer, J.~Gegelia, G.~Japaridze and S.~Scherer,
Int.\ J.\ Mod.\ Phys.\ A \textbf{27} (2012) 1250178 [arXiv:1211.1684
[hep-ph]].


\bibitem {lhc}
G.~Aad \textit{et al.} [ATLAS Collaboration],
Phys.\ Lett.\ B [arXiv:1207.7214 [hep-ex]]; S.~Chatrchyan \textit{et al.} [CMS
Collaboration],
Phys.\ Lett.\ B [arXiv:1207.7235 [hep-ex]].



\bibitem {nonlocal}

G.~V.~Efimov and M.~A.~Ivanov, \textquotedblleft The Quark confinement model
of hadrons,\textquotedblright\
\textit{Bristol, UK: IOP (1993) 177 p.}
Y.~V.~Burdanov, G.~V.~Efimov, S.~N.~Nedelko and S.~A.~Solunin,
Phys.\ Rev.\ D \textbf{54} (1996) 4483 [arXiv:hep-ph/9601344].
F.~Giacosa, T.~Gutsche and A.~Faessler,
Phys.\ Rev.\ C \textbf{71} (2005) 025202 [arXiv:hep-ph/0408085].




\bibitem {e38} F.~Giacosa and T.~Wolkanowski,
Mod.\ Phys.\ Lett.\ A {\bf 27} (2012) 1250229
[arXiv:1209.2332 [hep-ph]].


\bibitem {nonexp}F.~Giacosa, G.~Pagliara,
Mod.\ Phys.\ Lett.\ \textbf{A26 } (2011) 2247-2259. [arXiv:1005.4817
[hep-ph]].
G.~Pagliara, F.~Giacosa,
Acta Phys.\ Polon.\ Supp.\ \textbf{4 } (2011) 753-758. [arXiv:1108.2782
[hep-ph]].
F.~Giacosa, G.~Pagliara,
[arXiv:1110.1669 [nucl-th]].




\bibitem {duecan}F.~Giacosa,
Found.\ Phys.\ \textbf{42} (2012) 1262 [arXiv:1110.5923 [nucl-th]].


\bibitem {lupoder}
F.~Giacosa, G.~Pagliara,
Nucl.\ Phys.\ \textbf{A812 } (2008) 125-139. [arXiv:0804.1572 [hep-ph]].


\bibitem {kloe}
F.~Ambrosino \textit{et al.} [KLOE Collaboration],
Phys.\ Lett.\ B \textbf{634} (2006) 148 [hep-ex/0511031];
F.~Ambrosino \textit{et al.} [KLOE Collaboration],
Phys.\ Lett.\ B \textbf{681} (2009) 5 [arXiv:0904.2539 [hep-ex]].


\bibitem {pasca}P. Facchi and S. Pascazio, Quantum Probability and White Noise
Analysis XVII (2003) 222, quant-ph/0202127.










\bibitem {pdg}K. Nakamura \textit{et al}. (Particle Data Group), J. Phys. G
\textbf{37}, 075021 (2010).
\end{thebibliography}
\end{document}